\documentclass[pra, twocolumn,   showpacs]{revtex4}  
\usepackage{amsmath,amscd,amssymb,euscript,eufrak}
\usepackage{graphicx}
\usepackage{graphics}
\usepackage{epsfig}
\usepackage{ifthen}
\usepackage{verbatim}
\newcommand{\rmi}{\mathrm{i}}
\begin{document}
\title{Spin-orbit effects in armchair carbon nanotubes:\\analytical results}
\author{K.N. Pichugin$^{1}$, M. Pudlak$^{2}$,
and R.G. Nazmitdinov$^{3,4}$}
\affiliation{$^{1}$Kirensky Institute of Physics, 660036 Krasnoyarsk, Russia \\
$^{2}$Institute of Experimental Physics,
04001 Ko{\v s}ice, Slovak Republic\\
$^{3}$ Departament de F{\'\i}sica,
 Universitat de les Illes Balears, E-07122 Palma de Mallorca, Spain\\
$^{4}$ Bogoliubov Laboratory of Theoretical Physics,
Joint Institute for Nuclear Research, 141980 Dubna, Russia}
\begin{abstract}
Energy spectra and transport properties of armchair nanotubes with curvature induced spin-orbit interaction 
are investigated thoroughly. 
The spin-orbit interaction consists of two terms: 
the first one preserves the spin symmetry in rotating frame,
 while the second one breaks it. 
It is found that the both terms are equally important:  
i)at scattering on the potential step which mimics a long-range potential in the nanotubes;
ii)at transport via nanotube quantum dots.
It is shown that an armchair nanotube with the first spin-orbit term works as 
an ideal spin-filter, while the second term produces a parasitic inductance.
\end{abstract}
\date{\today}
\maketitle
\section{Introduction}
\label{intro}
Electronic and transport properties of carbon nanotubes are  highly topical 
subjects in mesoscopic physics (see for a review \cite{b,r1,r2,r3}) due to 
potential technological applications  in nano-electronics 
and optical devices \cite{sm,el}. Among various studies of physical properties of 
carbon nanotubes (CNTs), a detailed understanding of spin-orbit interaction  
is crucial for the interpretation of ongoing experiments as well as for 
future applications of the nanotubes in spintronics. 

 In general, the intrinsic (intraatomic) spin-orbit interaction in graphene 
is weak \cite{min}, since carbon atoms have zero nuclear spins,
and the hyperfine interaction of electron spins with nuclear spins is suppressed.
It makes a spin decoherence in such material to be 
weak as well, i.e., scattering due to disorder is supposed to be not important.
A full analysis of a spin-orbit interaction in CNTs requires, however, to consider  
the isospin degree of freedom present in the honeycomb carbon lattice.

According to a general wisdom, graphene being a zero-gap semiconductor has a band structure
described by a linear dispersion relation at low energy, similar 
to massless Dirac-Weyl fermions \cite{b,r4}. For a CNT the quantization condition 
leads, however, to metallic or semiconducting behavior, depending on chirality \cite{r1,r2}. 
The curved geometry may give rise to a band gap even for  metallic CNTs \cite{me}. 
Such a gap would allow to confine electrons, otherwise not possible due 
to the Klein paradox \cite{all}. 
However, a precise form of a spin-orbit interaction in a single-layer graphene
nanotube is still not well known.

A consistent approach to introduce the curvature-induced spin-orbit coupling (SOC)   
in the low-energy physics of graphene have been developed by Ando \cite{Ando} and by others 
\cite{Ent,Mart,Brat}.
Recent measurements in ultra clean CNTs \cite{Ku}, at various values of
the magnetic field, revealed the energy splitting which
can be associated with a spin-orbit coupling. Indeed,
the measured shifts are compatible with theoretical  
predictions \cite{Ando,Brat}. However, some features are left debatable. 
Evidently, removing 
the degeneracy between quantum levels,  the magnetic field 
generates new mechanisms as well, which obscure  effects 
related to a plain spin-orbit coupling 
(see, for example, discussion in \cite{Loss,Chico,Jeo,izum,val}).

It is noteworthy on the pivotal fact that
within the approach developed by Ando \cite{Ando}
one obtains two SOC terms: one preserves the spin symmetry 
in the rotating frame (see below), while the second one breaks this symmetry. 
In previous studies \cite{Ando,Brat} the role played by the second term was underestimated. 
The purpose of the present paper is twofold.
First, to consider consistently 
a full curvature-induced  spin-orbit coupling in an armchair nanotube within the approach 
suggested by Ando \cite{Ando}. Second, to show that 
the second term could play an important role in transport phenomena. 
In order to illuminate the role of interplay between both terms on electron transport,
we analyze the situation at zero magnetic field, removing all additional mechanisms 
related to the magnetic field.
In contrast to previous studies, we also provide analytical estimations for the
energy spectrum and transport coefficients for different cases 
(with and without the second term). Evidently, the analytical approach 
gives a fundamental insight into the nature of electronic and transport
properties of CNTs.

The structure of the paper is as follows. In Sec. II we derive an explicit formula 
for eigen spectrum of the Ando Hamiltonian with a full curvature-induced  spin-orbit 
coupling in an armchair nanotube. In Sec. III we discuss different symmetries associated with
the Hamiltonian and analyze the current operators. Sec. IV is devoted to the analysis
of scattering phenomena at the interface introduced by a potential step and to 
transport properties of carbon quantum dots at the preserved spin symmetry. In Sec. V, 
with the aid of results of Sec. IV, we discuss transport effects produced by a full 
curvature-induced  spin-orbit coupling in armchair nanotubes. Main conclusions are summarized
in Sec. VI.  Appendix provides technical details used for analytical solutions.

\section{The Model}
\label{sec:1}
\begin{figure}
\hspace{2cm}
\includegraphics[width=3.5cm,clip=]{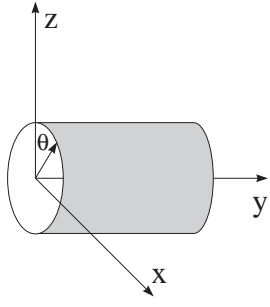}
\caption{The coordinate system for the carbon nanotube.} 
\label{fig_tube}
\end{figure}

Figure \ref{fig_tube} sketches a carbon nanotube and
a coordinate system with respect to 
the orientation axis of a carbon nanotube in our analysis.
The orbitals corresponding to the $\sigma$ bands of graphene are made by linear 
combinations of the $2s$, $2p_x$ $2p_y$ atomic orbitals, whereas 
the orbitals of the $\pi$ band are $p_z$ orbitals. 

Starting from the tight-binding model, with the aid of
${\bf k\cdot p}$ scheme in the vicinity of the Fermi energy ($E=0$) at $K$ and
$K^\prime$ points of the first Brillouin Zone, Ando has derived the effective mass 
Hamiltonian for electrons on curved surface with spin-orbit interaction 
(see details in \cite{Ando}).
We follow this approach and use the effective mass 
Hamiltonian as a starting point
of our analysis. This Hamiltonian can be expressed in the form of matrix-Hamiltonian 
equation
\begin{equation}
\label{1}
\hat{H}\Psi=
\left(\begin{array}{cc}0&\hat{f}\\
\hat{f}^{\dag}&0
\end{array}
\right)\left(
\begin{array}{c}
F^{K}_{A}\\F^{K}_{B}
\end{array}
\right)=E\left(
\begin{array}{c}F^{K}_{A}\\F^{K}_{B}\end{array}\right)\,,
\end{equation}
with the following definitions
\begin{eqnarray}
\label{1a}
&& \hat{f}=\gamma
(\hat{k}_{x}-\rmi\hat{k}_{y})+\rmi\frac{\delta\gamma'}{4R}\hat{\sigma}_x(\vec{r})-
\frac{2\delta \gamma p}{R}\hat{\sigma}_{y}\,,\\
&& \hat{k}_{x}=-\rmi\frac{\partial}{R\partial\theta}\,,
\hat{k}_{y}=-\rmi\frac{\partial}{\partial y}\,,\nonumber\\
&&\hat{\sigma}_x(\vec{r})=\hat{\sigma}_{x}\cos\theta -\hat{\sigma}_{z}\sin\theta\,.
\nonumber
\end{eqnarray}
Here, $\hat{\sigma}_{x,y,z}$ are standard Pauli matrices, 
and the spinors of two sub-lattices are
\begin{equation}
F^{K}_{A}=\left(\begin{array}{c}F^{K}_{A,\uparrow}\\F^{K}_{A,\downarrow}\end{array}\right)\,,\quad
F^{K}_{B}=\left(\begin{array}{c}F^{K}_{B,\uparrow}\\F^{K}_{B,\downarrow}\end{array}\right)\,.
\end{equation}
We preserve the definitions introduced by Ando \cite{Ando} for the following 
parameters: 
\begin{eqnarray}
\label{ggp}
&\gamma =-\sqrt{3}V_{pp}^{\pi}a/2\,,\nonumber\\
&\gamma'=\sqrt{3}(V_{pp}^{\sigma}-V_{pp}^{\pi})a/2\,,\\ 
&p=1-3\gamma'/8\gamma\,. \nonumber
\end{eqnarray}
Here, the quantities $V_{pp}^{\sigma}$ and $V_{pp}^{\pi}$ are the transfer 
integrals for $\sigma$ and $\pi$ orbitals, respectively in a flat 2D graphene, 
and $a$ is a lattice constant ($a=2.46$\AA).
The intrinsic source of the SOC 
$\delta = {\Delta}/(3\epsilon_{\pi\sigma})$ is defined by
\begin{equation}
\Delta = \rmi\frac{3\hbar}{4m_e^{2}c^{2}}\langle x_l|\frac{\partial
V}{\partial x} \hat{p}_{y}-\frac{\partial V}{\partial y} \hat{p}_{x}|y_l\rangle
\end{equation}
and $\epsilon_{\pi\sigma}=\epsilon_{2p}^{\pi}-\epsilon_{2p}^\sigma$, where
$V$ is the atomic potential.
The energy $\epsilon_{2p}^{\sigma}$ is the energy of
$\sigma$-orbitals which are localized between carbon atoms. The
energy $\epsilon_{2p}^{\pi}$ is the energy of $\pi$-orbitals which
are directed perpendicular to the nanotube surface. In our consideration 
$x_l$, $y_l$, and $z_l$ are local coordinates; $z_l$-axis is perpendicular 
to a graphene plane, and $y_l$-axis is lying along the tube symmetry axis.
\begin{figure*}
\includegraphics[width=4.5cm,clip=]{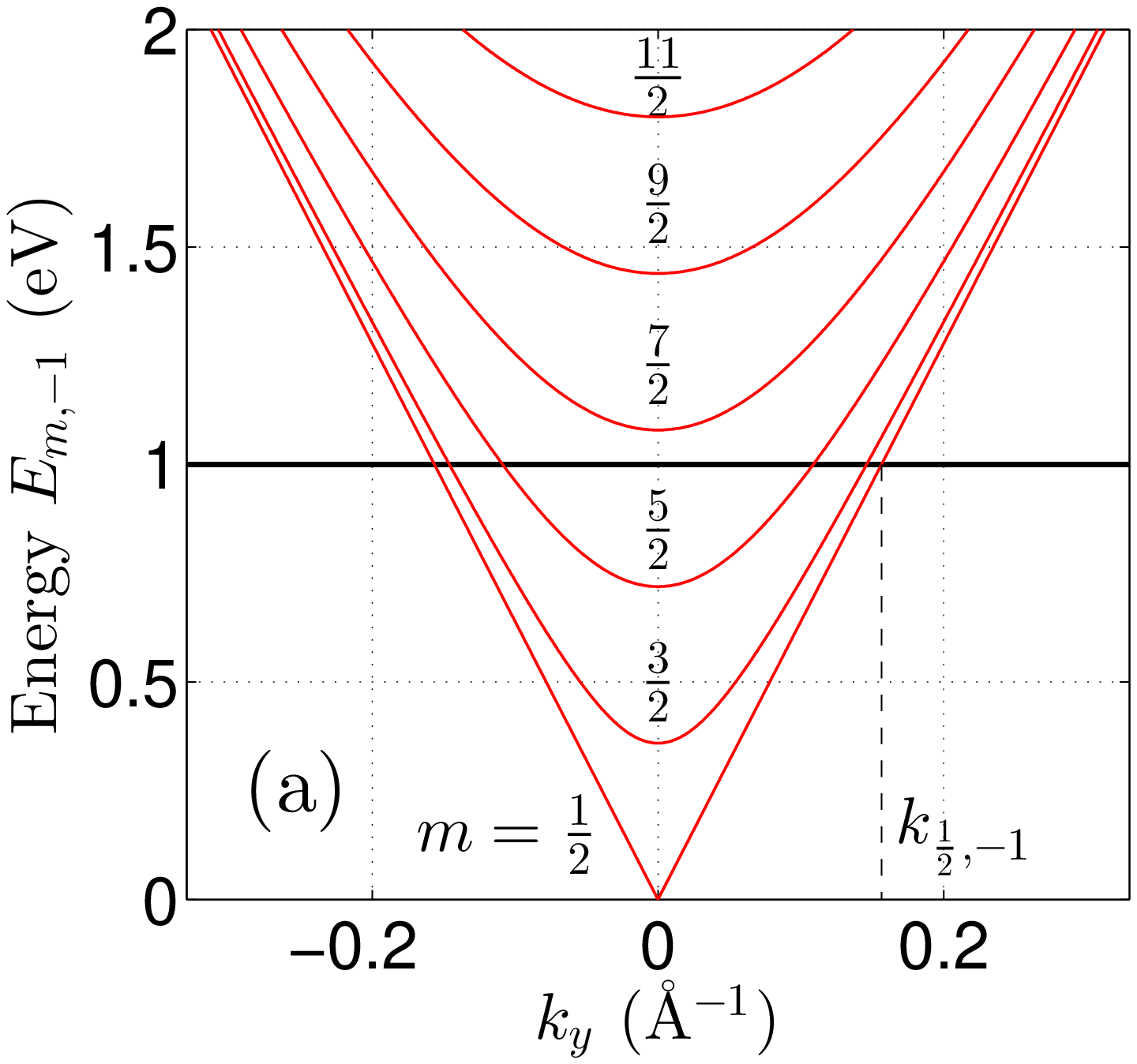}
\includegraphics[width=4.5cm,clip=]{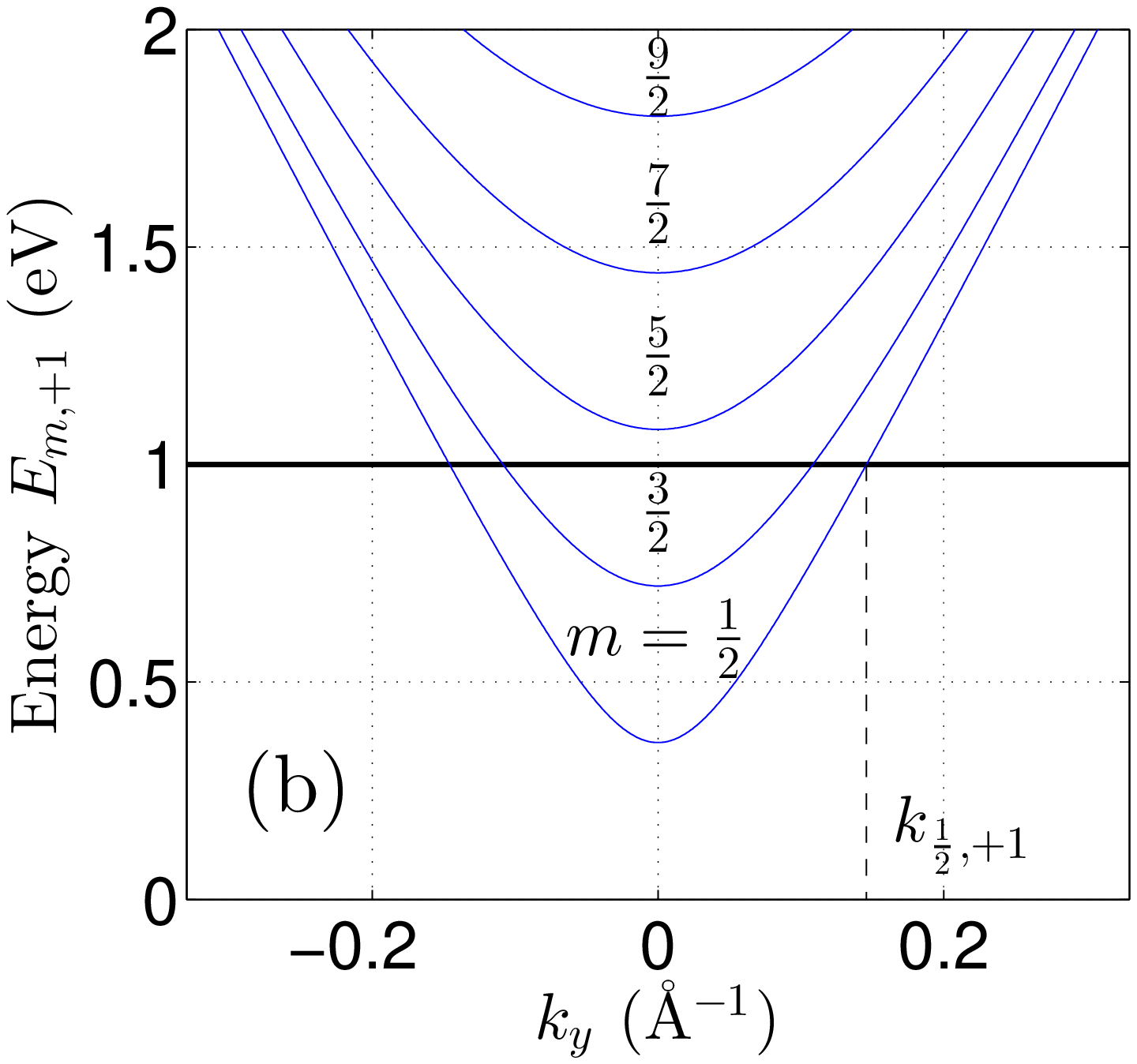}
\label{fig_E}
\hspace{1cm}
\vspace{-1cm}
\begin{minipage}[b]{7cm}
\caption{(Color online) Positive spectrum 
(see Eqs.(\ref{Epm_av}))
$E_{m,-1}$ (a) and $E_{m,+1}$ (b) as a function of the wave number $k_y$. The values of
$k_{m,+1}$ and $k_{m,-1}$ 
for $m=1/2$ at the energy $E_F=1$eV (solid horizontal line) are indicated by 
vertical dashed lines.
The parameters are $R=17.75$\AA, $\delta=0.01$, $p=0.1$,
$\gamma=\frac{9}{2}1.42$\AA eV, $\gamma'=\gamma \frac{8}{3}$, 
$\lambda_x=\frac{\gamma}{R}(1/2+2\delta p)=0.18$eV, 
$\lambda_y=\frac{\delta\gamma'}{4R}=0.0024$eV.
} 
\vspace{1cm}
\end{minipage}
\vspace{1cm}
\end{figure*}

Ando suggested to neglect a spin-orbit term proportional to $\sigma_x$, i.e.,
the term $\delta\gamma'/(4R)\hat{\sigma}_x(\vec{r})$ in the Hamiltonian (\ref{1}).
He assumed that a spin projection on the CNT symmetry axis (y-axis) is a conserved integral of motion.  
Based on the perturbative approach result,
he concluded that a spin mixing in the wave function due to this term
 is very small. Ando admitted, however,  that it may
couple states from  bands  with different spin quantum numbers and
lead to a small spin relaxation.
Although in  Ref.\cite{izum} a different basis was used 
to derive the effective Hamiltonian for a single wall CNT, the
term breaking its spin symmetry (a conservation of  ${\hat s_z}$-component) 
was obtained as well. 
Similar to  Ref.\cite{Ando},   
it was also suggested in Ref.\cite{izum} to neglect such a term.
In contrast, we consider all spin-orbit terms on equal footing,
since even a small perturbation brought about by the second term
breaks the fundamental  spin symmetry of the total Hamiltonian. 
In this case all three spin projections are not conserved in any system.
Different bands must be distinguished by the magnetic quantum number $m$ which
is a projection of the total angular momentum on the symmetry axis
of the CNT (see below).
As a result, this preserved fundamental symmetry allows to couple states with
different spins even inside one band.
We restrict our consideration by an armchair nanotube. 
In this case the SOC does not lead to the additional effect such as 
an electron-hole asymmetry \cite{izum,val}.

 To get rid of the $\theta$ dependence in the Hamiltonian (\ref{1}), we 
apply the transformation
\begin{equation}
\hat{H}'=\hat{U}\hat{H}\hat{U}^{-1}
\end{equation}
with the aid of the unitary operator 
$\hat{U}$ 
\begin{equation}
\label{U}
\hat{U}=\left(
\begin{array}{cc}
\exp({\rmi\frac{\theta}{2}}\hat{\sigma}_{y})&0\\
0&\exp({\rmi\frac{\theta}{2}}\hat{\sigma}_{y})
\end{array}\right)\,.
\end{equation}
As a result, we obtain the Hamiltonian in the transformed frame
\begin{eqnarray}
\label{2}
&& \hat{H}' = \hat{H}_{\it kin}+\hat{H}_{SOC}\,,\\
&& \hat{H}_{\it kin} = -\rmi\gamma \left(\hat{\tau}_y\otimes I\partial_y+
\hat{\tau}_x\otimes I
\frac{1}{R} \partial_\theta \right)\,,\\
&& \hat{H}_{SOC} = -\lambda_y\hat{\tau}_y\otimes\hat{\sigma}_x - 
\lambda_x\hat{\tau}_x\otimes\hat{\sigma}_y \,,
\end{eqnarray}
where $I$ is $2\times2$ unity matrix, and
\begin{equation}
\label{lxy}
\lambda_x=\gamma\left(1+4\delta p\right)/(2R)\,,\quad
\lambda_y=\delta\gamma^{\prime}/(4R)\,.
\end{equation}
We distinguish in the Hamiltonian (\ref{2}) the kinetic $\hat{H}_{\it kin}$ 
and the potential $\hat{H}_{SOC}$ terms.
Here, the operators $\hat{\tau}_{x,y,z}$ are the Pauli matrices
which act on the wave functions of A- and B-sub-lattices (a pseudo-spin space).
Note that the kinetic term couples the wave functions of A- and B-sub-lattices as well
as the potential term.

In our consideration, the curvature-induced spin-orbit coupling is described by
 two terms: $\lambda_x$ and $\lambda_y$. The term $\lambda_x$ depends on: 
i)values of the transfer integral $V_{pp}^{\pi}$ for $\pi$ orbitals; and 
ii)combined action produced by a product of 
the  {\it intrinsic} spin-orbit interaction 
and the difference between the transfer integrals $V_{pp}^{\pi}$, $V_{pp}^{\sigma}$  
for $\pi$  and $\sigma$ orbitals, respectively, in a flat 2D graphene.
The term $\lambda_y$ depends on the difference between 
the transfer integrals $V_{pp}^{\pi}$, $V_{pp}^{\sigma}$  
for $\pi$  and $\sigma$ orbitals, respectively, in a flat 2D graphene.
The both terms are inversely proportional to the tube radius, and
tend to zero at $R\to \infty$, i.e., in the limit of a 
flat graphene. For small nanotubes (small radius) we might expect, however, that 
effects produced by these terms come into particular prominence
in transport phenomena.

Herewith, for the sake of convenience, we use $\hbar=1$, if otherwise it will be
not mentioned. At $\lambda_y=0$, the  spin projection  $\hat{S}_y=\frac{1}{2}I\otimes \hat{\sigma}_y$ 
is a constant of motion, since it commutes with the Hamiltonian $[\hat{S}_y,\hat{H}']=0$. 
The term $\lambda_y\neq 0$ breaks this symmetry and yields a spin mixing. 
Due to an axial symmetry of the CNT, the projection of a total angular momentum
on the nanotube symmetry axis is always the integral of motion.
In our consideration, in the transformed system the integral of motion 
$\hat{J}_{y}$
\begin{equation}
\hat{J}_y=I\otimes\left(\hat{L}_y+\frac{\hat{\sigma}_y}{2}\right)=
I\otimes\left(-\rmi\partial_\theta+\frac{\hat{\sigma}_y}{2}\right),\; 
\left[\hat{H},\hat{J}_y\right]=0
\end{equation}
takes a simple form
\begin{equation}
\hat{J}_y\rightarrow 
\hat{J}_y'=\hat{U}\hat{J}_y\hat{U}^{-1}=
I\otimes\left(-\rmi\partial_\theta\right).
\end{equation}
In virtue of this fact, we consider the wave function in the form of plane waves
\begin{equation}
\label{4}
F'(\theta,y)=e^{\rmi m \theta}e^{\rmi k_y y}\Psi\,,
\end{equation}
where the wave function $\Psi$ is a four-component spinor.
The wave function (\ref{4}) defines 
the eigenvalues $m$ of the operator $\hat{J}_y$
\begin{equation}
\hat{J}_y'F'(\theta,y)=mF'(\theta,y)\,, m= \pm 1/2, \pm 3/2,....,
\end{equation}
while a quantum number $k_y$ is an  eigenvalue of the operator 
$\hat{k}'_y\equiv\hat{k}_y$  
\begin{equation}
\hat{k}_y'F'(\theta,y)=k_yF'(\theta,y)\,.
\end{equation}
Taking into account Eqs.(\ref{2},\ref{4}), we obtain our
Hamiltonian
\begin{eqnarray}
&\label{Hp}
\hat{H}'(m,k_y)=\\
&\left(\begin{array}{cccc}
0&0&t_m-\rmi t_y&\rmi(\lambda_y+\lambda_x)\\
0&0&\rmi(\lambda_y-\lambda_x)&t_m-\rmi t_y\\
t_m+\rmi t_y&-\rmi(\lambda_y-\lambda_x)&0&0\\
-\rmi(\lambda_y+\lambda_x)&t_m+\rmi t_y&0&0\\
\end{array}\right),
\nonumber
\end{eqnarray}
which acts on the spinor $\Psi$. Here we introduce the following 
definitions
\begin{equation}
\label{abbrev}
t_m=\frac{\gamma}{R}m,\; t_y=\gamma k_y.
\end{equation}

The solution of the eigenvalue problem of Hamiltonian (\ref{Hp}) yields the 
following four energies (see Appendix A, Eqs.(\ref{Epm_av}))
\begin{equation}
\label{Epm}
E^{(+)}=+E_{m,s}\,,E^{(-)}=-E_{m,s}\,,\quad s=\pm 1.
\end{equation}
We define all states with $E^{(+)}>0$ ($E^{(-)}<0$) as a particle (hole) states.
As was mentioned above, there is the electron-hole symmetry  
$|E^{(+)}|=|E^{(-)}|$.

The energy spectrum for a typical CNT as a function of the continuous 
variable $k_y$ and the quantized projection of the angular momentum $m$ 
is shown on Fig.2. For the sake of illustrations, in 
numerical calculations we use the following
parameters \cite{Ando}: 
$V_{pp}^{\pi}\sim-3$eV,  $V_{pp}^{\sigma}\sim 5$eV, $|p|\sim 0.1$,
the bond length $d=a/\sqrt{3}\approx$ 1.42\AA.
For a given $m$-value the Fermi 
energy $E_F$ provides  four possible values for $k_y$: $\pm k_{m,s}$.

The energy gap $\Delta E_{m,s}$ in the CNT is defined by a minimal distance 
between negative and positive parts of the spectrum (see Eqs.(\ref{Epm_av})) 
\begin{equation}
\label{gap}
\Delta E_{m,s}=2|E_{m,s}(k_y=0)|=2\left|\sqrt{t_m^2+\lambda_y^2}+s\lambda_x\right|\equiv2E^0_{m,s}\,.
\end{equation}
At $\lambda_y=0$ there is a minimal gap $\Delta E_{1/2,-1}=4\gamma\delta |p|/R$,
which coincides with the value obtained by Ando \cite{Ando}.
At $\lambda_x\neq0$ and $\lambda_y\neq0$ for $p>0$ the minimal gap becomes even lesser, 
while for $p<0$ it increases. Thus, the comparison of the gap (\ref{gap}) 
with experimental data would allow to fix the model parameters (\ref{ggp}).
The transport does not persist in the gap, since 
all eigen modes are evanescent ones. 
Evidently, when the spin-orbit interaction ($\delta=0$) is zero, one is faced
with a plain metallic CNT. 

\section{Symmetries and current operators}
\label{sec:2}

The Hamiltonian (\ref{2}) has several symmetries.
There is a particle-hole symmetry 
\begin{equation}
\hat{M}_a\hat{H}'\hat{M}_a^{-1}=-\hat{H}'\,,
\end{equation}
defined by the operator
\begin{equation}
\hat{M}_a=\hat{\tau}_z\otimes I\,.
\end{equation}
Therefore, energies for the eigenfunctions $\Psi$ and  $\hat{M}_a\Psi$ are equal
in value but opposite in sign.
There are two inversion operators
$M_\theta$, $M_y$, for $\theta$, $y$ coordinates, respectively,
\begin{equation}
\hat{M}_y=\hat{\tau}_y\otimes\hat{\sigma}_y\,,\quad \hat{M}_\theta= \hat{\tau}_y\otimes\hat{\sigma}_x,
\end{equation}
with properties
\begin{eqnarray}
&&\hat{M}_y\hat{H}'(\hat{k}_y,\hat{J}_y)\hat{M}_y^{-1}=\hat{H}'(-\hat{k}_y,\hat{J}_y)\,,\\
&&\hat{M}_\theta\hat{H}'(\hat{k}_y,\hat{J}_y)\hat{M}_\theta^{-1}=\hat{H}'(\hat{k}_y,-\hat{J}_y)\,.
\end{eqnarray}
These transformations connect the eigenfunctions with opposite quantum numbers
$k_y$ and $m$.

In virtue of the conservation law for the current ${\bf j =j_y+j_\theta}$
\begin{equation}
\frac{\partial}{\partial t}|\Psi|^2+\nabla {\bf j}=0\,,
\end{equation}
we obtain a longitudinal and an orbital current operators 
\begin{equation}
\label{J}
\hat{j}_y=\gamma \hat{\tau}_y\otimes I\,,\quad \hat{j}_\theta=\gamma \hat{\tau}_x\otimes I.
\end{equation}
The same expressions can be obtained from the equation of motion
\begin{equation}
\hat{v}=\hat{\dot{r}} = \rmi [\hat{H}',\hat{r}]\,.
\end{equation}

For fixed quantum numbers $(k_y,m)$, at  $E>0$ the current moves 
in a direction opposite to one of the current at $E<0$. 
This fact follows from the symmetry relation
\begin{equation}
\hat{M}^{-1}_a \hat{j}_{y,\theta}\hat{M}_a=-\hat{j}_{y,\theta}\,.
\end{equation}

The expectation values of the $\theta$ ($y$)-component of the current 
calculated by means of the eigenfunctions $\Psi$ and $M_\theta\Psi$ ($M_y\Psi$) are of opposite sign. 
This result  follows from the identities 
\begin{eqnarray}
&&\hat{M}^{-1}_{\theta} \hat{j}_{\theta}\hat{M}_{\theta}=-\hat{j}_{\theta}\,,\quad
\hat{M}^{-1}_{y} \hat{j}_{\theta}\hat{M}_{y}=\hat{j}_{\theta}\,,\\
&&\hat{M}^{-1}_{\theta} \hat{j}_{y}\hat{M}_{\theta}=\hat{j}_{y}\,,\quad
\hat{M}^{-1}_{y} \hat{j}_{y}\hat{M}_{y}=-\hat{j}_{y}\,.
\end{eqnarray}

With the aid of the relations
\begin{eqnarray}
&&\hat{M}^{-1}_\theta \hat{S}_{y,z}\hat{M}_\theta=-\hat{S}_{y,z}\,,\\
&&\hat{M}^{-1}_\theta \hat{S}_x\hat{M}_\theta=\hat{S}_x\,,\\
&&\hat{M}^{-1}_y \hat{S}_{x,z}\hat{M}_y=-\hat{S}_{x,z}\,,\\
&&\hat{M}^{-1}_y \hat{S}_y\hat{M}_y=\hat{S}_y\,,
\end{eqnarray}
it could be shown that the expectation values of spin projections onto the local $y,z$ 
($x,z$)-axes for the eigenfunctions $\Psi$ and $M_\theta\Psi$ ($M_y\Psi$) 
have also opposite signs.

\section{Analytical results at $\lambda_y=0$}
\label{ly0}

To understand how the full SOC affects the system properties,
we consider first only $\lambda_x\neq 0$, $\lambda_y=0$.
As discussed above, in this case  
the operator $\hat{S}_y=\frac{1}{2}I\otimes\hat{\sigma}_y$
is an integral of motion. Therefore, we transform our Hamiltonian to the 
frame where the operator $\hat{S}_y$ has a diagonal form
\begin{equation}
\hat{\cal V}^{-1}\hat{S}_y\hat{\cal V}=\frac{1}{2}\hat{\sigma}_z\otimes I\,.
\end{equation}
Here, the transformation $\hat{\cal V}$, defined as
\begin{equation}
\hat{\cal V}=
\left(\exp\left(\rmi\frac{\pi}{4}\hat{\sigma}_x\right)\otimes I\right)\hat{P}_{23}\,,
\hat{P}_{23}=\left(
\begin{array}{cccc}
 1 & 0 & 0 & 0 \\
 0 & 0 & 1 & 0 \\
 0 & 1 & 0 & 0 \\
 0 & 0 & 0 & 1
\end{array}
\right)\,,
\end{equation}
consists of a rotation on angle $\pi/2$ around $x$-axis and
the permutation $P_{23}$. This permutation collects spin up components of the  A- and B-sub-lattices 
in the upper part of the spinor $\Psi$.
In virtue of this transformation, the Hamiltonian (\ref{Hp}) gains
a block-diagonal structure 

\begin{eqnarray}
&\hat{\cal H}=\hat{\cal V}^{-1}\hat{H}'\hat{\cal V}=
\left(
\begin{array}{cccc}
 0 & a_{-} & 0 & 0 \\
 a_{+} & 0 & 0 & 0 \\
 0 & 0 & 0 & b_{-} \\
 0 & 0 & b_{+} & 0
\end{array}
\right)\,,
\\
&a_{\pm}=(t_m-\lambda_x)\pm\rmi t_y\,,\quad b_{\pm}=(t_m+\lambda_x)\pm\rmi t_y\,.
\nonumber 
\end{eqnarray}

One obtains obvious four eigenvalues and 
eigenvectors
\begin{eqnarray}
\label{EV_l0}
& E^{(\pm)}=\pm E_{m,s},
 \quad E_{m,s}=\sqrt{(t_m+s\lambda_x)^2+t_y^2}\,,\\ 
\label{vm}
& V_{m,+1}=\frac{1}{\sqrt{2}}(0,0,1,z_{m,+1})^T\,,\\
&V_{m,-1}=\frac{1}{\sqrt{2}} (1,z_{m,-1},0,0)^T\,,\\
\label{zm}
& \quad z_{m,s}=((t_m+s\lambda_x)+\rmi t_y)/E\,.
\end{eqnarray}

\subsection{Scattering on a potential step}
\label{ssps}

\begin{figure}
\hspace{2cm}
\includegraphics[width=5cm,clip=]{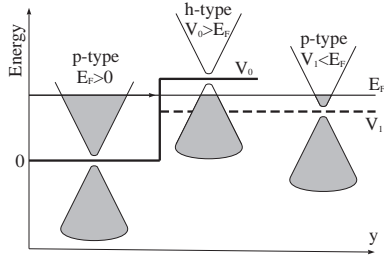}
\caption{A schematic illustration of the scattering process 
on a potential step.} 
\label{fig_trans}
\end{figure}

We consider first the scattering
at the interface introduced by a potential step
of the height $V_0$ as sketched in Fig.\ref{fig_trans}.
On the one hand, the step is assumed to be smooth on the length scale 
of a graphene unit cell (an inverse Brillouin momentum $2\pi/K$). 
Therefore, it does not induce the intervalley ($K\to K'$) scattering.  
On the other hand, it is assumed to be sharp on the Fermi length scale 
($\lambda\sim 1/k_F$). Since $\theta$ and $y$ are independent variables, 
such a potential conserves the angular momentum projection $m$. 

Depending on the sign of $E_F$ (incoming particle) and 
the sign of $E_F-V_0$ (outgoing particle) there are
four types of transmission through the step: p-p, p-h, h-p, h-h. 
We denote particle by symbol "p", while a hole by a symbol "h". 
For the sake of illustration, p-p and p-h transmissions are schematically 
shown on Fig.\ref{fig_trans}.

Transmission (reflection) of electrons incoming from the left is controlled 
by partial coefficients. Namely, we have $|t^q_{q'}|^2(|r^q_{q'}|^2)$ describing 
transmission (reflection) probability from the
left states with a set of quantum numbers $q=\{m,s\}$ to the right 
states with $q'=\{m',s'\}$.
Transmission (reflection) probabilities 
are defined as squares of the scattering-wave-function amplitudes 
which satisfy the continuity of wave functions. 
Evidently, since spin is a good quantum number as well as a quantum number $m$, 
the transmission (reflection) 
probabilities, responsible for the spin-flip process, are absent at 
$\lambda_y=0$: $|t^{m,s}_{m,s'}|^2=|r^{m,s}_{m,s'}|^2=0$ for $s\neq s'$.

\begin{figure*}
\includegraphics[width=4.5cm,clip=]{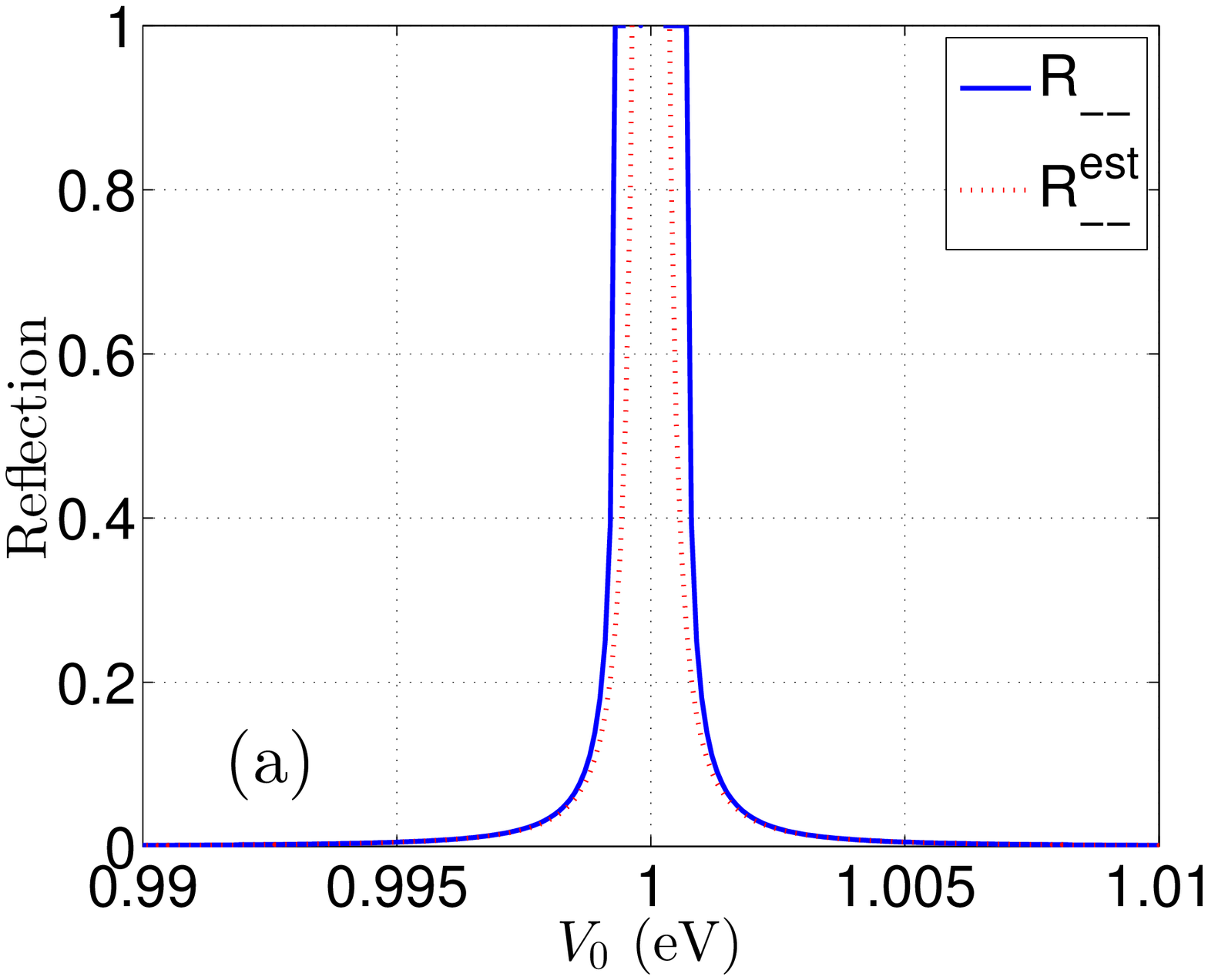}
\includegraphics[width=4.5cm,clip=]{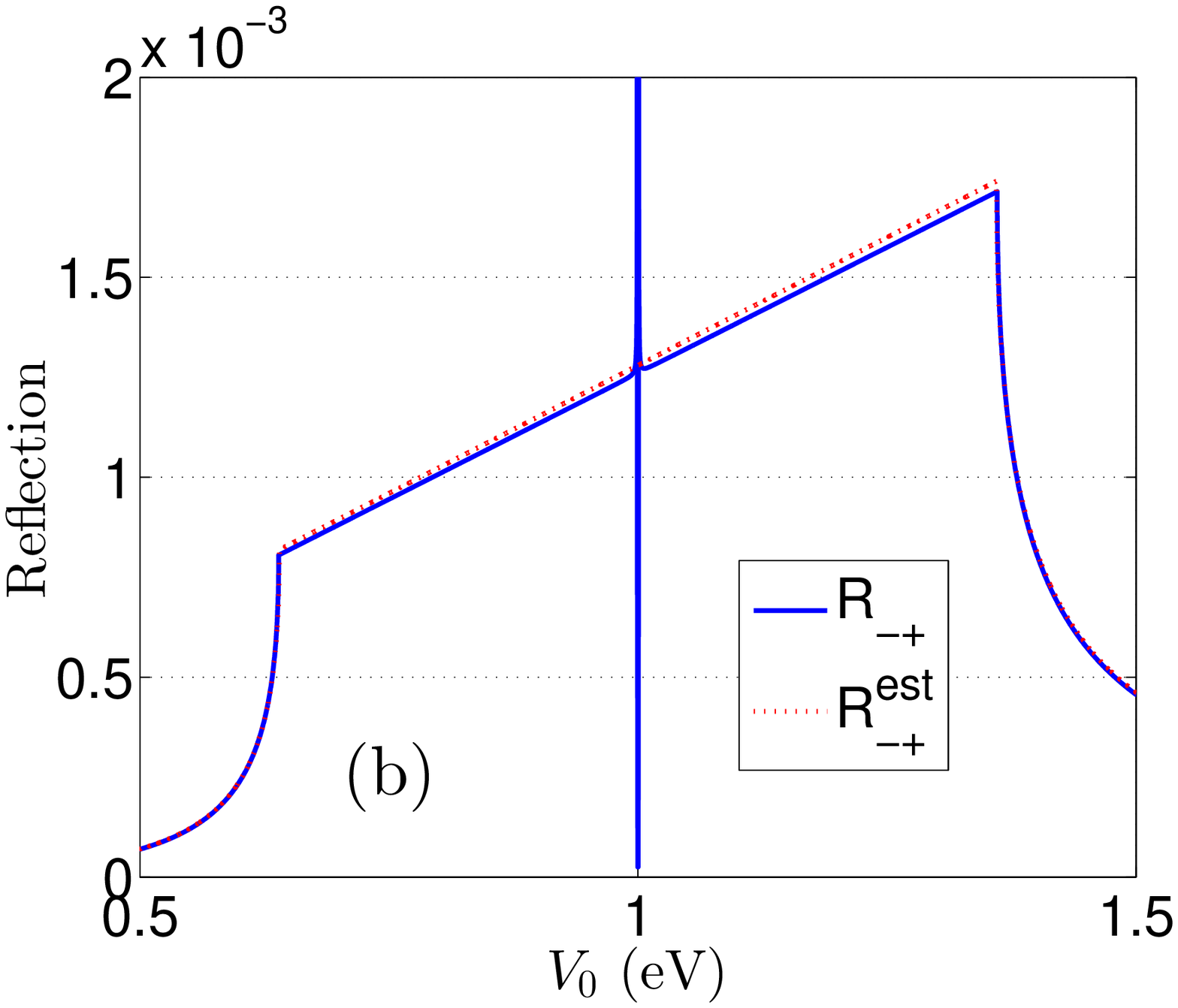}
\label{fig_exp}
\hspace{1cm}
\begin{minipage}[b]{7.5cm}
\caption{(Color online) Reflection probabilities on the potential step
for the state $m=1/2$, $s=-1$:
analytical estimation (dotted line, red) and numerical results (solid line, blue) 
as a function of the potential height $V_0$.
(a) $\lambda_y=0$ : a direct reflection $R_{--}^{\rm est}(V_0)=|r_{m,s}^{m,s}(V_0)|^2$
is described by Eq.(\ref{rnn_l0});
(b) $\lambda_y\neq 0$: a spin-flip reflection $R_{-+}^{\rm est}(V_0)=
|r_{m,-s}^{m,s}(V_0)|^2$ is described by Eq.(\ref{rnn}).
$E_F=1$eV and the other parameters are the same as 
in Fig.2.
} 
\vspace{0.5cm}
\end{minipage}
\vspace{1cm}
\end{figure*}
Matching the eigenfunctions with the same values of the angular momentum 
and spin projections on the left and right sides of the potential step, we obtain 
the following equation
\begin{equation}
\label{matchS_l0}
\frac{\left[
\left(
\begin{array}{c}
 1 \\ Z_q
\end{array}
\right)+
r_q^q
\left(
\begin{array}{c}
 1 \\ Z_q^{-1}
\end{array}
\right)\right]}
{\sqrt{|\gamma\rmi(z_q^*-z_q)|}}
=\frac{t_q^q}{\sqrt{|\gamma\rmi({\tilde z}_q^*-{\tilde z}_q)|}}
\left(
\begin{array}{c}
 1 \\ {\tilde Z}_q
\end{array}
\right).
\end{equation}
Here $Z_q=(z_q)^{{\rm sign}(E_F)}$ (${\tilde Z}_q=({\tilde z_q})^{{\rm sign}(E_F-V_0)}$) 
are defined by Eq.(\ref{zm}),
$z_q^*$ (${\tilde z}_q^*$) are complex conjugate, and 
the energy $E=E_F$ ($E=E_F-V_0$) before (after) the step. 
The wave functions are normalized to have a unit current flow along the $y$-axis.
The solution of Eq.(\ref{matchS_l0}) defines the reflection and transmission 
coefficients
\begin{eqnarray}
\label{rt_l0}
&&r_q^q=(Z_q-{\tilde Z}_q)/({\tilde Z}_q-Z_q^{-1})\,,\\
\label{tr_10}
&&t_q^q=(Z_q-Z_q^{-1})/({\tilde Z}_q-Z_q^{-1})
\sqrt{\left|\frac{{\tilde z}_q-{\tilde z}_q^*}{z_q-z_q^*}\right|}\,.
\end{eqnarray}

In order to reveal the effect of the SOC 
let us consider the case of quantum numbers 
$q_n:\{m=\mp\frac{1}{2},s=\pm 1\}$. It corresponds to
a normal incident direction of electrons on the potential 
step for the CNT without the SOC and $\delta=0$. 
Note that the variable $z_q$ (${\tilde z}_q$) (see Eq.(\ref{zm})) 
depends on the term $t_m+s\lambda_x$. In our case this term 
transforms to the form
\begin{equation}
t_m+s\lambda_x=\frac{\gamma}{R}\left(m+s/2+2sp\delta \right)
\Rightarrow \pm\frac{\gamma}{R} 2p\delta \,,
\end{equation}
which depends on a small parameter $\delta$.
The Taylor series expansion of the reflection $r_{q_n}^{q_n}$ 
(see Eq.(\ref{rt_l0})) 
over this parameter $\delta$ enables to us to define
the first nonzero term (omitting unimportant phase factor)
\begin{equation}
\label{rnn_l0}
r_{q_n}^{q_n}=\delta p\frac{\gamma}{R}\frac{V_0}{E_F(E_F-V_0)}+O(\delta^2)\,.
\end{equation}
The estimation (\ref{rnn_l0}) describes remarkably well numerical results for
the scattering until $E_F$ or $E_F-V_0$ is close 
to the gap (see Fig.4a). Thus, at $|E_F|\gg E_{q_n}^0$ and
$|E_F-V_0|\gg E_{q_n}^0$, where $E_{q_n}^0=2|\delta p\gamma|/R$ 
is of order $10^{-4}$eV for the typical parameters of CNTs, 
there is a weak backward scattering  defined by Eq. (\ref{rnn_l0}). 

To gain a better insight into scattering phenomena we study the current.
At $\lambda_y=0$ the expectation value of the 
longitudinal current $\langle\hat{j}_y\rangle$ is determined by the quantum number
$k_y$. The effect of the SOC is visible in
the expectation value of the orbital ($\theta$-) component 
of the current (see Appendix A, Eq.(\ref{J_av}))
\begin{equation}
\label{Jt_l0}
\langle\hat{j}_\theta\rangle_q=\frac{\gamma}{E}\left(t_m+s\lambda_x\right)\,,
\end{equation}
which depends on the spin-orbit term $\lambda_x$.
With the aid of Eqs.(\ref{lxy}), (\ref{abbrev}), one obtains that
the orbital current is always nonzero. Thus, we can have a persistent current
without a magnetic field. In order to understand this result
let us assume that there is a magnetic field along the symmetry axis $y$. 
It results in the Aharonov-Bohm magnetic flux passing through 
the CNT cross section, which yields the modified quantum number 
$m'=m+\phi/\phi_0$ ($\phi$ -- magnetic flux, $\phi_0$ -- magnetic flux quanta). 
Evidently, with the aid of the magnetic field along the symmetry axis y 
one can suppress the orbital current for any value of the quantum number $m$ 
(without the Zeeman splitting). 
Taking into account Eqs.(\ref{lxy}), (\ref{abbrev}), (\ref{Jt_l0}),  we obtain
the condition for a zero orbital current for a magnetic quantum number $m$
\begin{equation}
\label{CMF}
t_{m'}=-s\lambda_x\Rightarrow \phi/\phi_0=-(\frac{s}{2}+m)-2s\delta p\,.
\end{equation}
In particular, for the set of quantum numbers  $q_n=\{m=\mp\frac{1}{2}, s=\pm 1\}$
we obtain $\langle\hat{j}_\theta\rangle_q=0$ at the magnetic flux
$\phi/\phi_0=-2s\delta p$ which compensates the SOC term $\lambda_x$ 
at $m=-1/2$. 
Note that at the condition (\ref{CMF}) (and $\lambda_y=0$) 
the gap (\ref{gap}) vanishes as well for any $m$.

Thus, the SOC works as an "effective magnetic field", 
responsible for the orbital motion and, therefore, for
the weak backward scattering (\ref{rnn_l0}).
In the absence of Zeeman splitting the applied magnetic field 
$\phi/\phi_0=-2s\delta p$ leads to zero backward scattering in all orders.

\subsection{Quantum dot}
\label{ssqd}

\begin{figure}[ht]
\hspace{2cm}
\includegraphics[width=5cm,clip=]{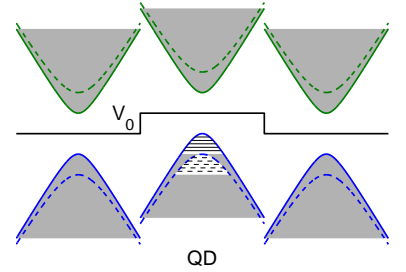}
\caption{(Color online) A schematic picture of quantum dot in CNT. Dashed lines correspond to
different sets of conserved quantum numbers.} 
\label{fig_QDtube}
\end{figure}

The gap in CNTs (due to the SOC) opens a possibility  
to create a nanotube quantum dot (QD), by confining particles 
in a quantum well with a potential $V_0(\theta(y)-\theta(L-y))$
(where $\theta(x)$ is a Heaviside step functions), 
as illustrated in Fig.\ref{fig_QDtube}.
We recall that in the gap there are evanescent modes only.

The QD energies are located within the gap $-E_q^0<E<E_q^0$ (see Eq. (\ref{gap})).
Different sets of quantum numbers $q=\{m,s\}$ determine the full spectrum of the QD.
Matching the eigenfunctions (\ref{vm}) at the $y=0$ 
and $y=L$ we obtain the following equations
\begin{eqnarray}
\label{matchQD_l0}
&r_l\left(
\begin{array}{c}
 1 \\ z_q^{-1}
\end{array}
\right)=
a
\left(
\begin{array}{c}
 1 \\ {\tilde z}_q
\end{array}
\right)+
b
\left(
\begin{array}{c}
 1 \\ {\tilde z}_q^{-1}
\end{array}
\right)\,,\\
&r_r\left(
\begin{array}{c}
 1 \\ z_q
\end{array}
\right)=
ae^{\rmi k_q(E) L}
\left(
\begin{array}{c}
 1 \\ {\tilde z}_q
\end{array}
\right)+
be^{-\rmi k_q(E) L}
\left(
\begin{array}{c}
 1 \\ {\tilde z}_q^{-1}
\end{array}
\right)\!\!\!,
\end{eqnarray}
where 
\begin{eqnarray}
\label{QD_zq_l0}
&z_q(E)=\left[(t_m+s\lambda_x)-\kappa_q(E)\gamma\right]/E\,,\nonumber\\
&\kappa_q(E)=\sqrt{(t_m+s\lambda_x)^2-E^2}/\gamma\,,\nonumber\\
&{\tilde z}_q(E)=\left[(t_m+s\lambda_x)+\rmi k_q(E)\gamma\right]/(E-V_0)\,,\\
&k_q(E)=\sqrt{(E-V_0)^2-(t_m+s\lambda_x)^2}/\gamma\,.\nonumber
\end{eqnarray}
These equations could be written in the matrix form
\begin{equation}
\label{QD_l0}
\left(
\begin{array}{cccc}
 -1 & 1 & 1 & 0 \\
 -z_q^{-1} & {\tilde z}_q & {\tilde z}_q^{-1} & 0 \\
 0 & e^{\rmi k_q(E) L} & e^{-\rmi k_q(E) L} & -1 \\
 0 & {\tilde z}_qe^{\rmi k_q(E) L} & {\tilde z}_q^{-1}e^{-\rmi k_q(E) L} & -z_q
\end{array}
\right)\,
\left(
\begin{array}{c}
 r_l \\ a \\ b \\ r_r 
\end{array}
\right)=0
\end{equation}

\begin{figure}
\vspace{-1cm}
\hspace{1cm}
\includegraphics[width=6cm]{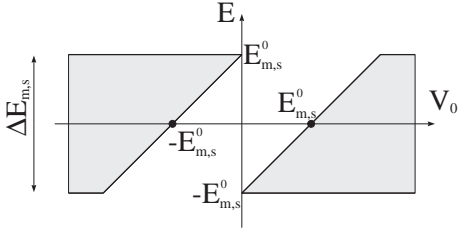}
\caption{Grey regions correspond to the area where discrete spectrum is possible.
} 
\label{fig_QD_area}
\end{figure}

Evidently, the solutions exist, if the determinant of Eq.(\ref{QD_l0}) is zero.
This requirement yields the transcendental equation which defines 
the eigen spectrum $E_q^n$ of the QD:
\begin{equation}
\label{spectr_l0}
\exp(2\rmi k_q(E_q^n) L)=\left(\frac{z_q(E_q^n) {\tilde z}_q(E_q^n)-1}{z_q(E_q^n)
-{\tilde z}_q(E_q^n)}\right)^2\,,
\quad n=1,2,3...
\end{equation}
Eq.(\ref{spectr_l0}) can be transformed to the form 
\begin{equation}
\label{spectrm_l0}
\tan(k_q(E_q^n) L)=\left(\frac{\gamma^2k_q(E_q^n) \kappa_q(E_q^n)}{E_q^n 
(E_q^n-V_0)-(t_m+s\lambda_x)^2}\right).
\end{equation}
As it was mentioned above, the QD spectrum is defined 
in the energy window $-E_q^0<E_q^n<E_q^0$.
Its boundaries are shown on Fig. \ref{fig_QD_area}.
Thus, Eq.(\ref{spectrm_l0}) corresponds to the case $\lambda_y=0$, when 
there is no spin scattering. In other words, in such QDs an electron
with spin up cannot scatter into a state with spin down and vise versa,
without any additional mechanism.

In order to gain a better insight into the properties of the QD spectrum, 
let us consider two limiting cases: a) $\gamma k_q(E_q^n)\gg|t_m+s\lambda_x|$ ; and
b) $\gamma k_q(E_q^n)\ll|t_m+s\lambda_x|$. 
In case a), with the aid of Eq.(\ref{QD_zq_l0}) we have 
\begin{equation}
\gamma k_q(E_q^n)\approx |E_q^n-V_0|\gg t_m+s\lambda_x\,,\quad z_q(E_q^n)\approx \pm\rmi\,.
\end{equation}
As a result, Eq. (\ref{spectr_l0}) transforms to the form
\begin{equation}
\exp\left(2\rmi\frac{|E_q^n-V_0|}{\gamma} L\right)\approx
\left(\frac{\pm\rmi {\tilde z}_q(E_q^n)-1}{\pm\rmi-{\tilde z}_q(E_q^n)}\right)^2\approx const.\,.
\end{equation}
At the condition $E_q^{n+1}-E_q^n \ll V_0$ it 
yields an equidistant spectrum, similar to the one of 
the harmonic oscillator potential:
\begin{equation}
E^{n+1}_q-E^n_q\approx \frac{\gamma\pi}{L}.
\end{equation}
In case b), at small $k_q(E_q^n)$  we obtain
\begin{eqnarray}
&\gamma k^n_q(E_q^n)=\sqrt{(E_q^n-V_0)^2 -(t_m+s\lambda_x)^2}\ll t_m+
s\lambda_x\,,\nonumber\\
&\quad z_q(E_q^n)\approx \pm 1\,.
\end{eqnarray}
As a result, Eq. (\ref{spectr_l0}) takes the form  
\begin{equation}
\exp\left(2\rmi k^n_q(E_q^n) L\right)\approx 1\,.
\end{equation}
At the condition $E_q^{n+1}-E_q^n \ll t_m+s\lambda_x$ it
defines the spectrum similar to the one of the quantum well potential:
\begin{equation}
E^{n+1}_q-E^n_q\approx \left(\frac{\gamma\pi}{L}\right)^2\frac{2n+1}{2(t_m+s\lambda_x)}\,.
\end{equation}
Both limits (low-and high-energies) should be  fulfilled for long nanotubes.

\section{General case}

\subsection{Scattering on a potential step}
\label{sspsg}

Analytical solutions of the eigenvalue problem for the Hamiltonian (\ref{Hp}) 
with nonzero $\lambda_x$ and $\lambda_y$ are presented in Appendix A. 
With the aid of these results we reconsider the scattering
at the interface introduced by a potential step
of the height $V_0$ (Fig.\ref{fig_trans}).
Unfortunately, analytical expressions for the general case are too cumbersome, and
we present mostly numerical results. We use the same typical values for 
graphene nanotubes (as in the previous section)  
to demonstrate a general tendency.

\begin{figure}
\hspace{1cm}
\includegraphics[width=5cm,clip=]{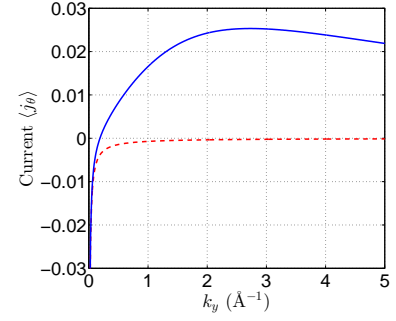}
\caption{(Color online) Current 
$\langle\hat{j}_\theta\rangle_q$ 
(see Eq.(\ref{Jt})) as a function of the wave number $k_y$ for the eigenstate with 
$q=\{m=1/2,s=-1\}$ at $\lambda_x\neq 0$ for: $\lambda_y\neq 0$ (solid line) 
and $\lambda_y=0$ (dashed line).
The other parameters are the same as in 
Fig.2. At $k_y\approx 0.19$\AA$^{-1}$ the 
current is zero. 
} 
\label{fig_jt}
\end{figure}

The expectation value of the orbital ($\theta$) current component 
(see Appendix A, Eq.(\ref{J_av}))
\begin{equation}
\label{Jt}
\langle\hat{j}_\theta\rangle_{m,s}=
\frac{\gamma t_m}{E}\left(1+s
\frac{\lambda_x^2}{\sqrt{\lambda_x^2(t_m^2+\lambda_y^2)+t_y^2\lambda_y^2}}
\right)
\end{equation}
essentially depends on the spin-orbit term 
$\lambda_y\neq 0$, ignored in literature (see Fig.\ref{fig_jt}).

The orbital current  becomes zero at
\begin{equation}
|m|=1/2\,, s=-1\,, 
|t_y|=\frac{\lambda_x}{\lambda_y}\sqrt{\lambda_x^2-\lambda_y^2-t_m^2}\,,
\end{equation}
which corresponds to the energy $E=\pm1.199$eV 
for the parameters listed in the caption of Fig.2. 
To avoid the additional back scattering due to non-normal
incidence of electrons on the potential step, 
we use this energy and the quantum number $m=1/2$ for incoming electron 
to trace the transmission and reflection events as a function  of $V_0$ 
(Fig.8). 
For energies $E-V_0\leq 0.5$eV the conductance 
\begin{equation}
G=\frac{e^2}{h}(|t^{m,+1}_{m,+1}|^2+|t^{m,-1}_{m,-1}|^2+|t^{m,+1}_{m,-1}|^2+|t^{m,-1}_{m,+1}|^2)
\end{equation}
is dominated by the transmission without spin-flip (see Fig.8a,c),
i.e., by the probabilities $T_{++}=|t^{m,+1}_{m,+1}|^2$ and 
$T_{--}=|t^{m,-1}_{m,-1}|^2$, while the reflection is suppressed. 
At $E-V_0>0.5$eV the SOC gives rise to the direct reflection $R_{++}=|r^{m,+1}_{m,+1}|^2$
(without spin-flip), which grows rather rapidly.
Within the energy gap $-E_{m,+1}^0<E_F-V_0<-E_{m,-1}^0$ and 
$E_{m,-1}^0<E_F-V_0<E_{m,+1}^0$ ($E_{1/2,+1}^0\approx 0.36$eV) 
the transmission probability $T_{--}\approx 1$,
while the transmission probability $T_{++}$ is almost suppressed, since 
the reflection probability $R_{++}\approx 1$.
Although the probability $T_{--}$ is 
a dominant process, the probability $T_{+-}=|t^{m,+1}_{m,-1}|^2$ produces a parasitic 
loss of this dominance due to $\lambda_y$ term in the SOC. Note that the 
transmission probability $T_{-+}=|t^{m,-1}_{m,+1}|^2$ is completely suppressed in this energy window.

\begin{figure}
\includegraphics[width=4cm,clip=]{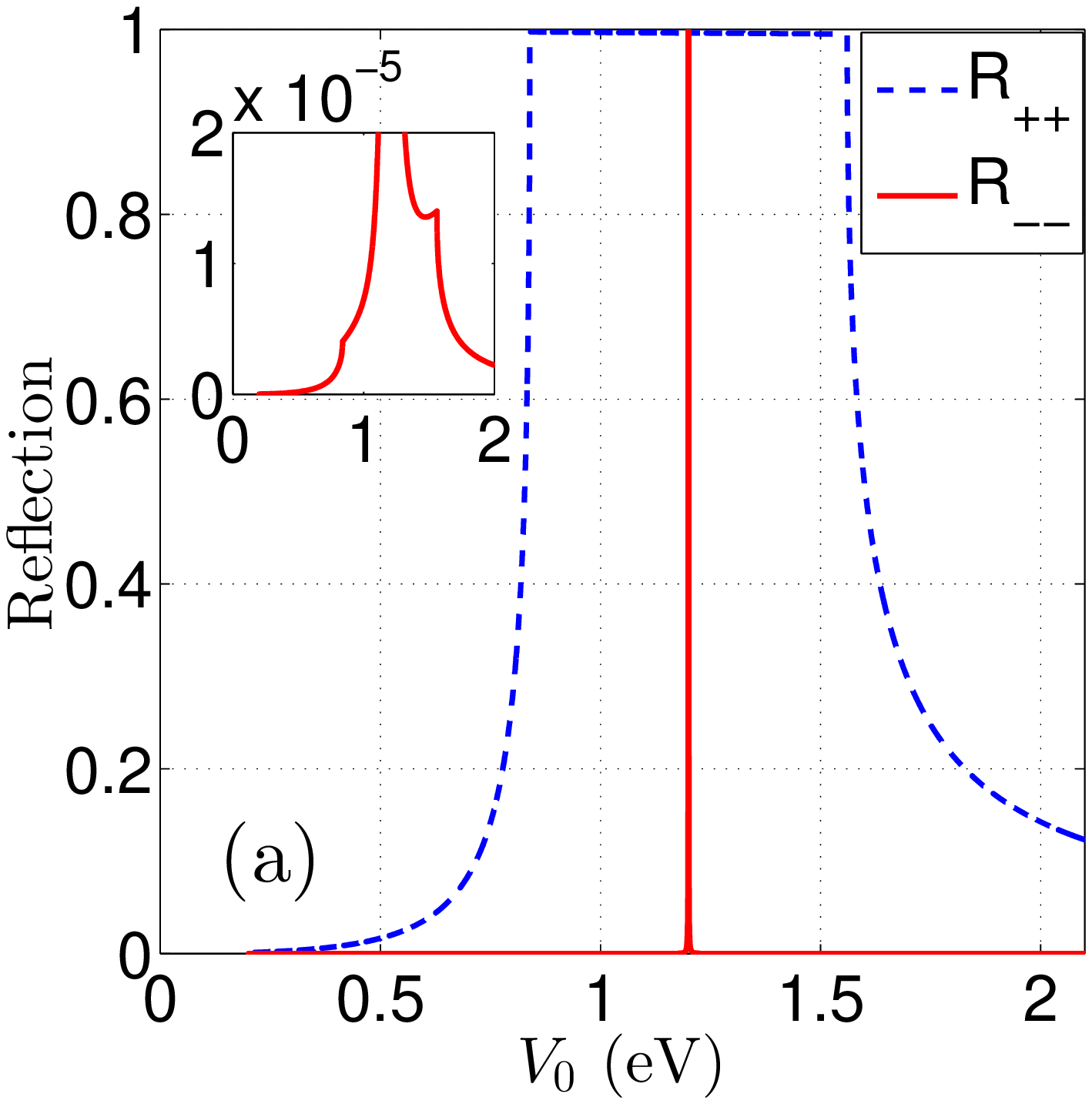}
\includegraphics[width=4cm,clip=]{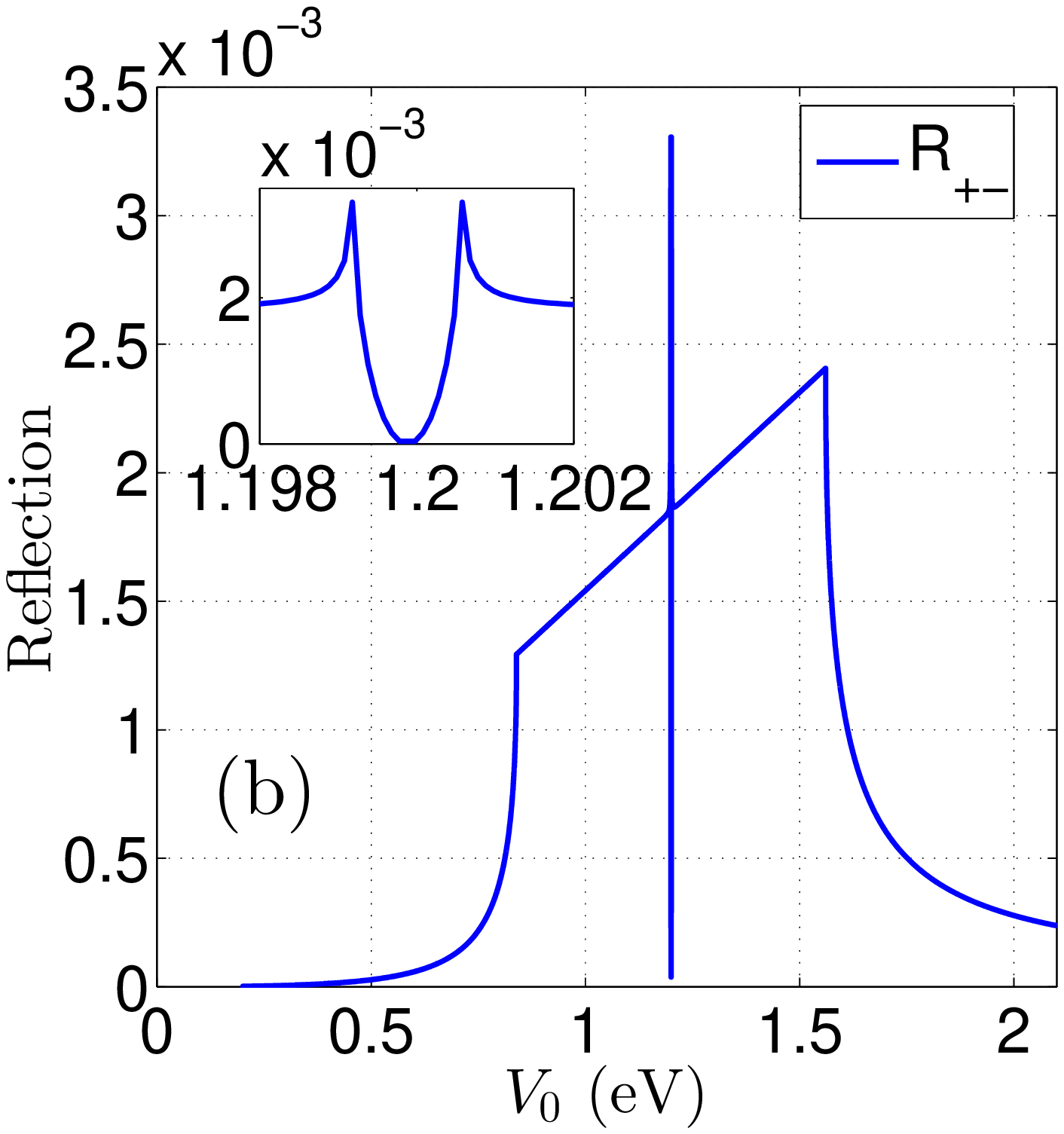}\\
\includegraphics[width=4cm,clip=]{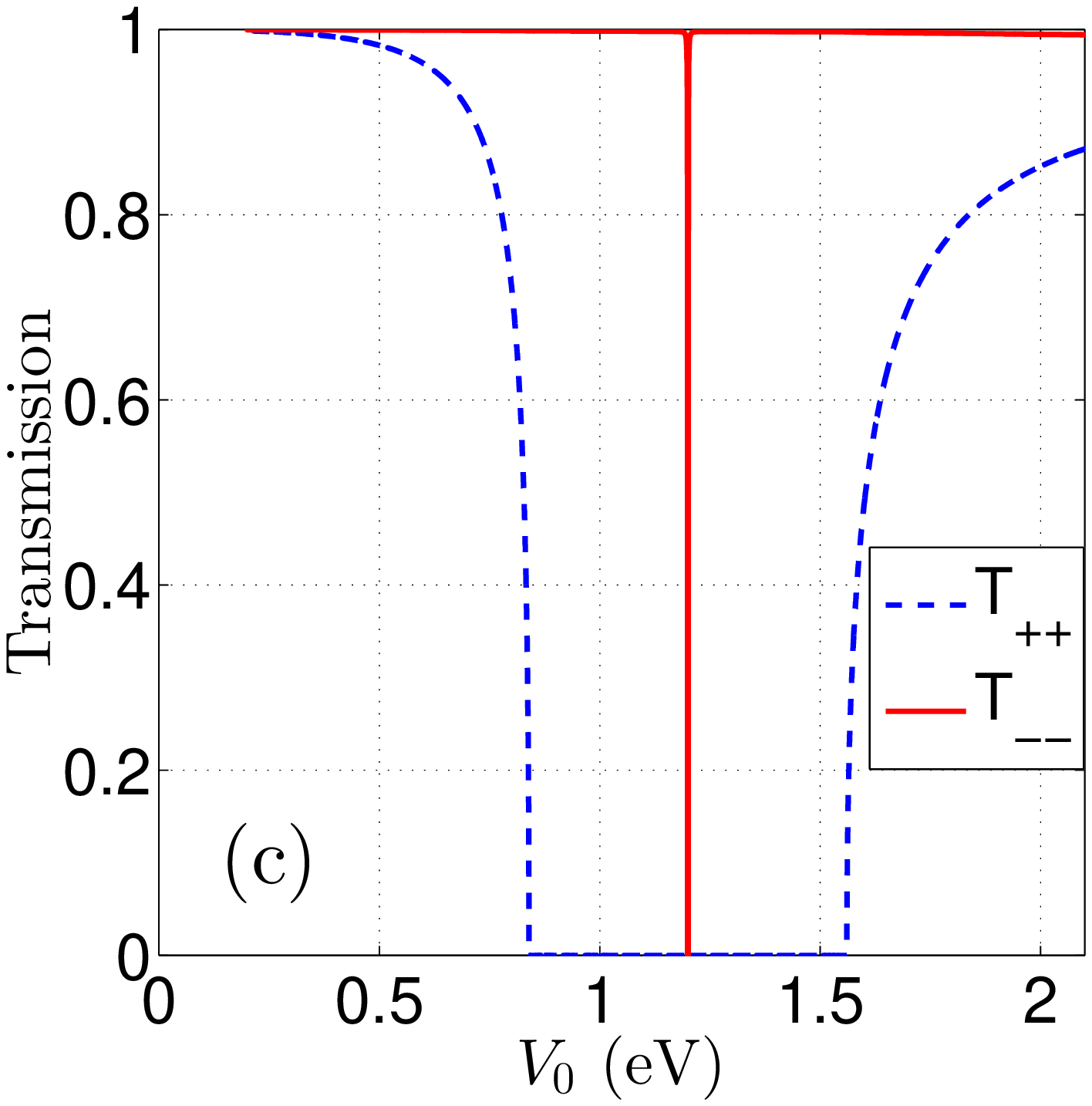}
\hspace{-0.2cm}
\includegraphics[width=3.8cm,clip=]{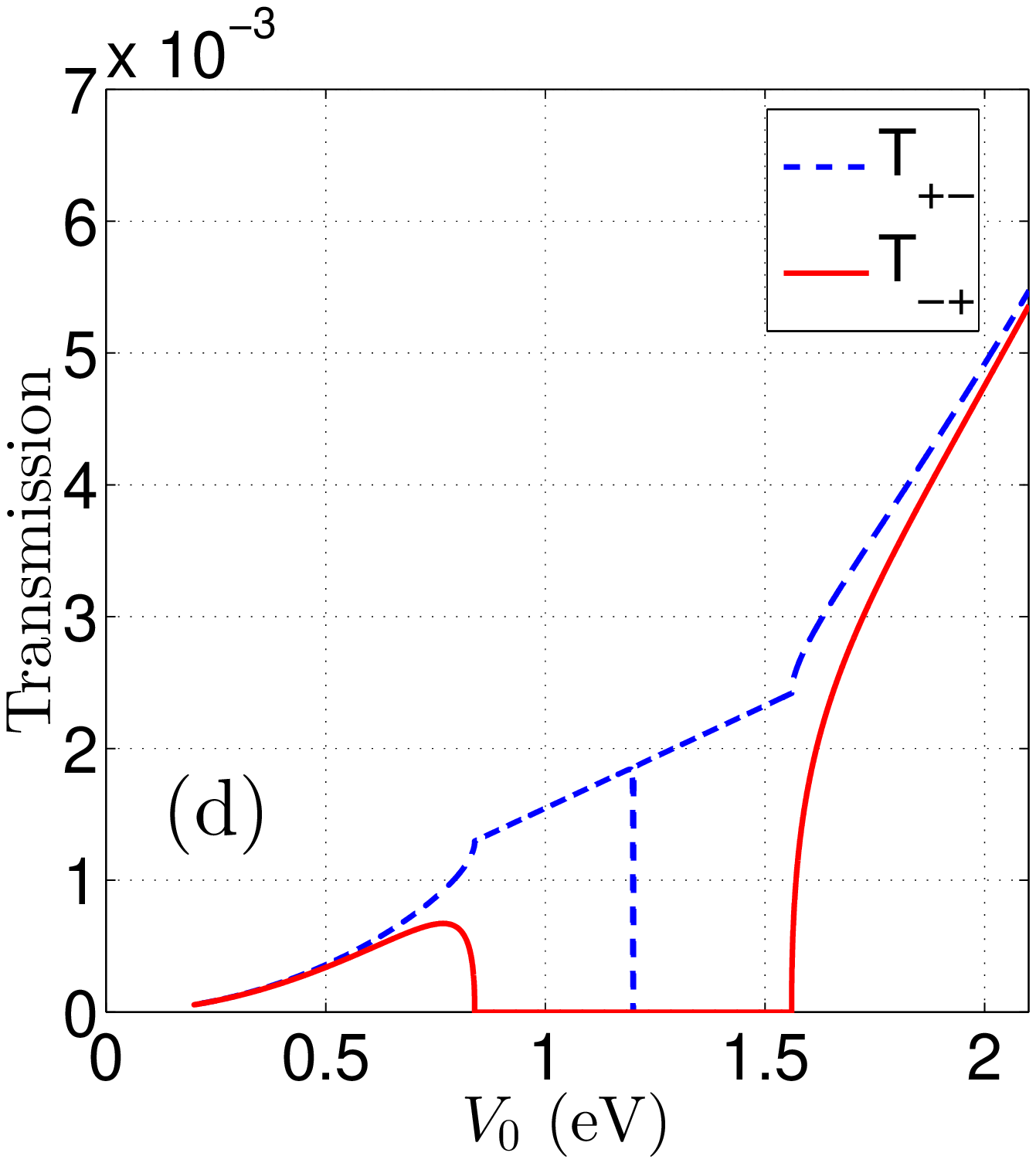}
\vspace{0.5cm}
\begin{minipage}[b]{8cm}
\caption{(Color online) Dependence of reflection (a), (b) and transmission (c), (d)
probabilities for normal incidence at 
$m=1/2$, $E_F=1.199$eV on the height of the potential step $V_0$ (eV). 
The parameters are the same as in Fig.2.
} 
\end{minipage}
\vspace{-0.5cm}
\label{fig_RV}
\end{figure}
This mechanism resembles in appearance to the one considered for 
one-dimensional electron system formed in semiconductor heterostructures 
showing strong Rashba spin-orbit interaction in the presence 
of weak magnetic field \cite{SebaPRL2003}. In our case, the spin-filter effect is 
brought about by a relatively weak curvature-induced SOC which creates the 
"effective magnetic field". It is noteworthy that at $\lambda_y=0$ the filter 
would be even more efficient due to the absence of the inter-channel scattering.

There is a full reflection zone which corresponds to the SOC's induced gap
of the width $\Delta E_{1/2,-1}=2*0.7meV\approx 16^{\circ}K$ for chosen 
parameters. In this energy interval the evanescent modes exist only.
In contrast to the case considered in the previous section ($\lambda_y=0$), 
there is a mixing of spin components. 
In general, the back scattering
\begin{equation}
G_{bs}=\frac{e^2}{h}(|r^{m,+1}_{m,+1}|^2+|r^{m,-1}_{m,-1}|^2+|r^{m,+1}_{m,-1}|^2+|r^{m,-1}_{m,+1}|^2)
\end{equation}
being small increases  on two order in magnitude in the presence of
the SOC induced by $\lambda_y$ term (compare the inserts on Fig.8a,b).
For completeness, we consider reflection for 
$q_n=\{m=\pm\frac{1}{2},s=\mp 1\}$, which corresponds to a normal incidence 
for the CNT without the SOC. The expansion of reflection amplitudes over the parameter 
$\delta$, which is responsible for the {\it intrinsic} graphene spin-orbit interaction, leads to the 
results
\begin{eqnarray}
\label{rnn}
&r^{m,s}_{m,s}=\delta p\frac{\gamma}{R}\frac{V_0}{E_{in}E_{out}}+O(\delta^2)\,,\\
\label{rnn1}
&r^{m,s}_{m,-s}=\frac{\lambda_y}{2t_m^2}V_0\times
\left[(E_{in}^2-4t_m^2)(E_{in}-2 |t_m|)^2\right]^{1/4}\times\nonumber\\
&\times {\cal Z}_1/{\cal Z}_2+O(\delta^2)\,,\\
&{\cal Z}_1={\rm sign}(E_{out})\sqrt{E_{out}^2-4t_m^2}-(E_{out}-2|t_m|)\,,\\
&{\cal Z}_2=(E_{out}-2 |t_m|)\sqrt{E_{in}^2-4t_m^2}+\nonumber\\
&+{\rm sign}(E_{in}){\rm sign}(E_{out})(E_{in}-2 |t_m|)\sqrt{E_{out}^2-4t_m^2}\,.
\end{eqnarray}
Here, we use the following notations:
$E_{in}=E_F, E_{out}=E_F-V_0$, omitting unimportant phase factors. 
The direct reflection amplitude (\ref{rnn}) is described 
in the lowest order by the same formula (\ref{rnn_l0}) as for the case  $\lambda_y=0$. 
Indeed, the term $\lambda_y$ contributes to the 
direct scattering only in the second order with respect to the strength $\delta$. 
Its contribution is, therefore, negligible in a direct scattering.
The spin-flip reflection appears, however, due to $\lambda_y$ term solely. 
Eqs.(\ref{rnn}),(\ref{rnn1}), reproduce remarkably well the complex behavior of reflection 
displayed on Fig.8b (for a comparison see Fig.4b). 
\begin{figure}
\includegraphics[width=4cm,clip=]{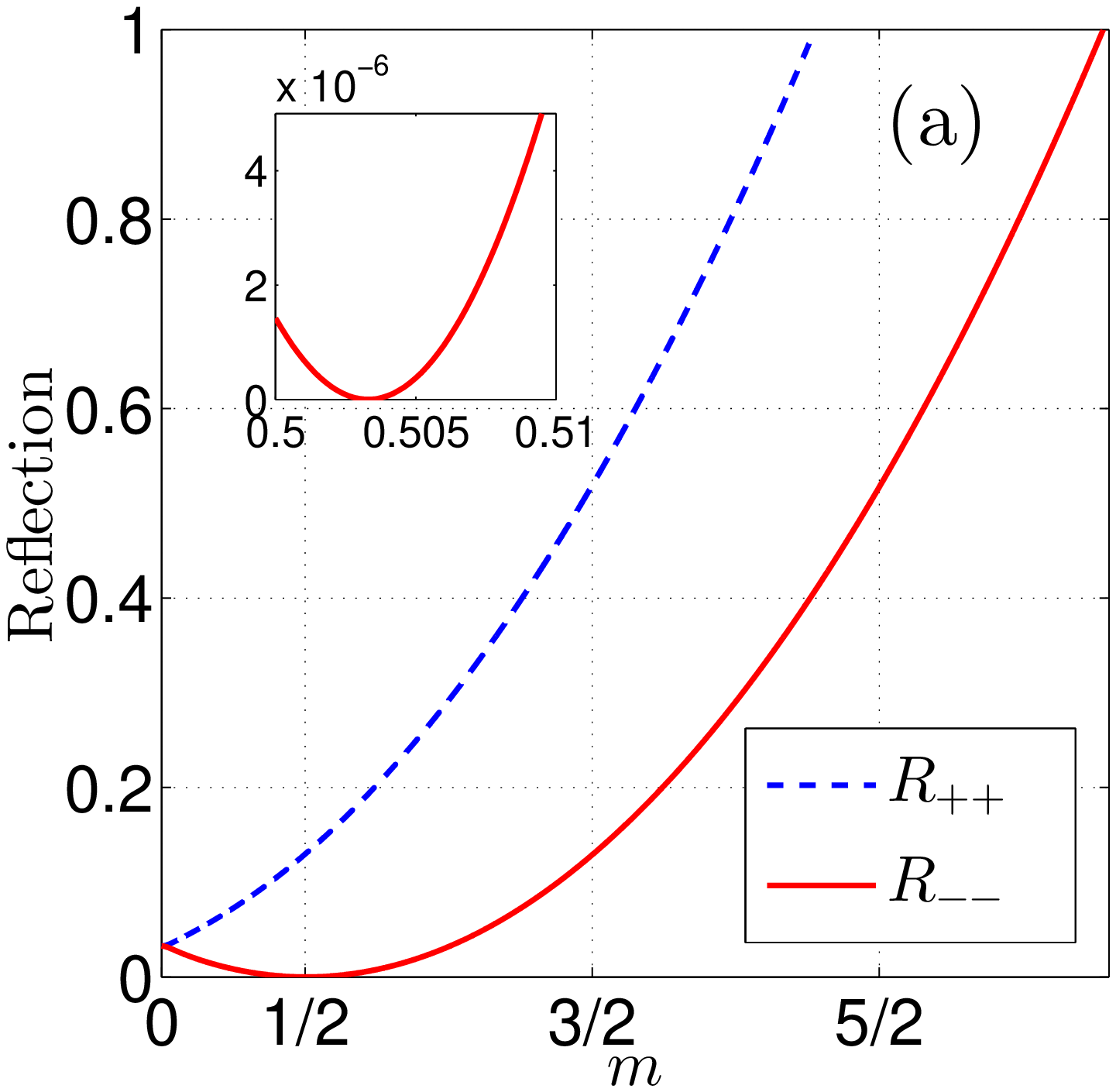}
\includegraphics[width=4cm,clip=]{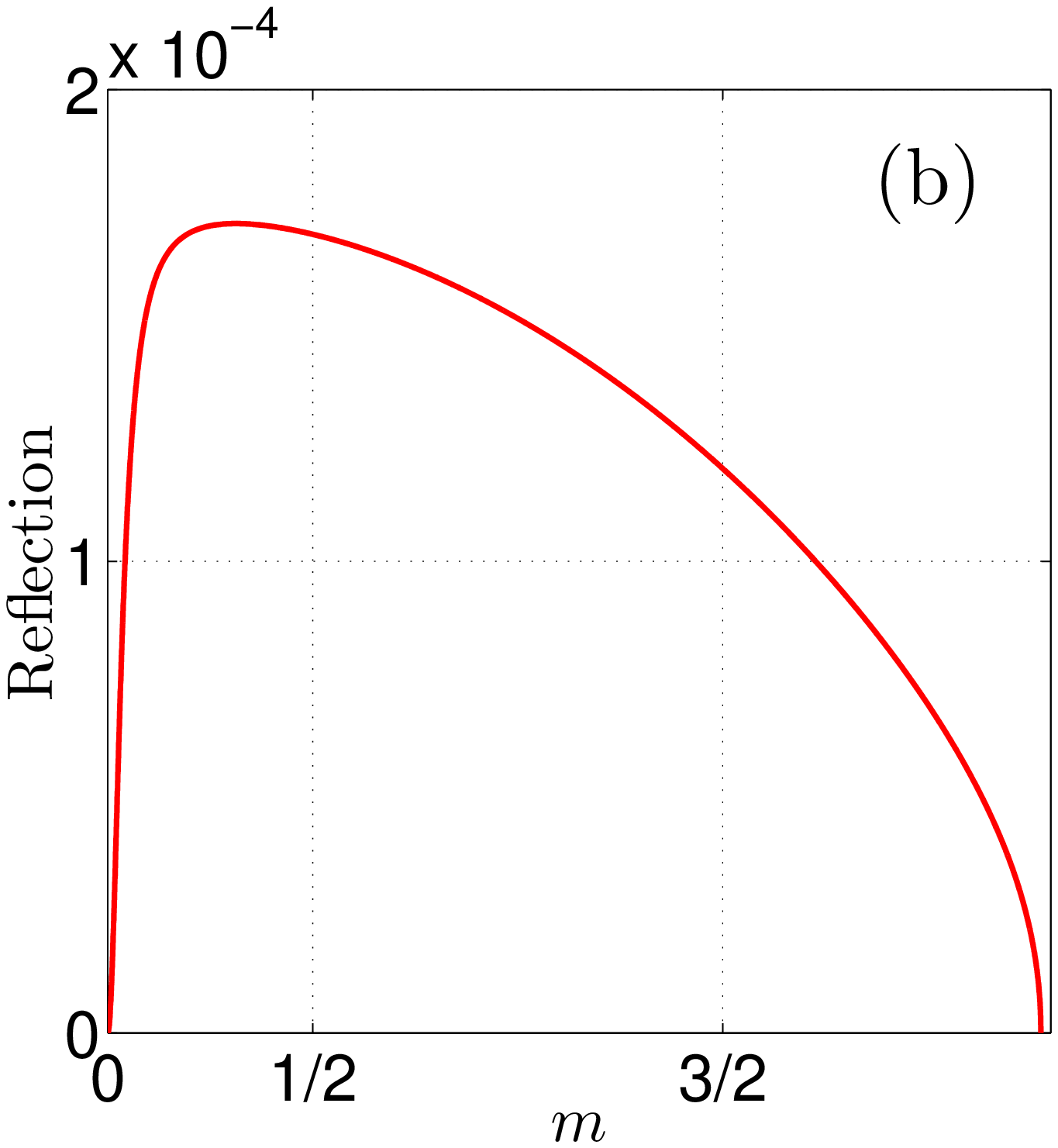}
\caption{(Color online) Reflection probabilities as a function of the angular 
momentum projection $m$.
The energy of incoming (outgoing) electron is $E_F=1$eV($E_F-V_0=-1$eV).
The panel (a) displays the direct reflection probabilities: 
$R_{--}=|r^{m,-1}_{m,-1}|^2$ (solid, red) and $R_{++}=|r^{m,+1}_{m,+1}|^2$ 
(dashed, blue). The panel (b) displays the reflection probabilities with the spin-flip:
$R_{-+}=|r^{m,-1}_{m,+1}|^2=|r^{m,+1}_{m,-1}|^2$.
The parameters are the same as in Fig.2.
} 
\label{fig_Rj}
\end{figure}
The reflection probabilities for different  $m$ are displayed
on Fig.\ref{fig_Rj}. The direct reflection (without the spin-flip) grows 
rapidly with the increase of the magnetic quantum number $m$ (see Fig.\ref{fig_Rj}a). 
In contrast, the spin-flip reflection probabilities tends to zero with the increase 
of the magnetic quantum number (see Fig.\ref{fig_Rj}b).
In comparison with the case  $\lambda_y=0$, the minimum in 
the reflection probability $|r^{m,-1}_{m,-1}|^2$ is slightly shifted from 
$m'=1/2+2p\delta$ (see the insert in Fig.\ref{fig_Rj}a)). 
In contrast to the case with $\lambda_y=0$, 
the direct reflection probability $|r^{m,-1}_{m,-1}|^2$ can not be turned to zero by 
the Aharonov-Bohm magnetic flux  alone, created by the magnetic field along the 
nanotube symmetry ($y$-) axis. The larger is the quantum number $m$ the smaller 
is the transmission probability.

The maximal magnetic quantum number $m$, which corresponds to
a complete reflection, could be determined from the condition
that the longitudinal current $\langle\hat{j}_y\rangle_{m,s}=0$ (see Appendix A).
This condition requires $t_y=\gamma k_y=0$ after the potential step. 
Note that the quantum number $s$ is not conserved at $\lambda_y\neq0$.
As a result, one obtains with the aid of Eqs.(\ref{Epm_av})
for the energy of outgoing electron the condition
\begin{eqnarray}
&&t_y=0\Rightarrow |E_F-V_0|=E_{m,s}^0\equiv |E_{m,s}(k_y=0)|\,,\\
&&E_{m,s}(k_y=0)=\sqrt{t_m^2+\lambda_y^2}+s\lambda_x\,,\quad s=\pm 1\,.
\end{eqnarray} 
At fixed parameters $\{E_F,V_0,\lambda_{x,y},R\}$, one defines the boundaries
\begin{equation}
\label{eq2}
C_{\pm} = t_m^2=\left(\frac{\gamma}{R}m\right)^2=(|E_F-V_0|\mp\lambda_x)^2-\lambda_y^2.
\end{equation}
Note that $C_->C_+$, and, therefore, the maximal $m=M$ is determined as
\begin{equation}
M=\frac{R}{\gamma}\sqrt{C_-}\,.
\end{equation}
For all $|m|\geq M$ the transmition probability 
$T_{s-}\equiv0$ for $s=\pm1$.
In particular, for our choice of parameters the reflection 
probability $R_{--}=|r^{m,-1}_{m,-1}|^2=1$ at $M\approx 7/2$ 
(compare with Fig.\ref{fig_Rj}a).

From now on we can define the critical angle of the complete reflection.
In virtue of  results from Appendix A we obtain for the critical angle  
\begin{eqnarray}
\label{tg}
&&\tan(\phi)=\frac{\langle j_\theta\rangle_q}{\langle j_y\rangle_q}=
\frac{t_M}{t_y}\left(1+\frac{\lambda_x^2-\lambda_y^2}{W}\right),\\
&&W=\sqrt{\lambda_y^2E_F^2+t_M^2(\lambda_x^2-\lambda_y^2)}\,,
\end{eqnarray}
where the critical value of the magnetic quantum number $m$ is related to
the variable $t_M$
\begin{equation}
t_M=\sqrt{C_-}=\sqrt{(|E_F-V_0|+\lambda_x)^2-\lambda_y^2}\,.
\end{equation}
We use $s=+1$ in Eq.(\ref{tg}), since $t_y(s=+1)<t_y(s=-1)$, where $t_y$ is defined
by Eq.(\ref{kpm_av})
\begin{equation}
t_y=\gamma k_s=\sqrt{E_F^2+\lambda_y^2-t_M^2-\lambda_x^2-
2sW}\,.
\end{equation}
These equations might provide some hint on the
contribution of different SOC terms at experimental measurements of the critical angle.

\subsection{Basic features of a quantum dot}
\label{ssqdg}

\begin{figure}
\hspace{1cm}
\includegraphics[width=6cm,clip=]{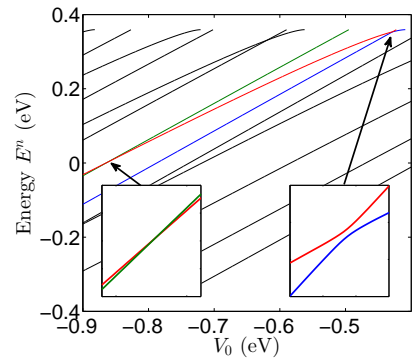}
\caption{(Color online) The QD energy spectrum as a function of applied potential $V_0$.
The levels with a quantum number $m=3/2$ are 
displayed for the nanotube with a length $L=100${\AA}. The other
parameters are the same as in Fig.\protect\ref{fig_E}. 
At different values of the applied potential, levels either cross 
(left insert) or repel (right insert) each other.
} 
\label{QDs}
\end{figure}
The energy spectrum of a QD is defined by the minimal gap for 
a given angular momentum $m$:
$-E_{m,-1}^0<E^n_m<E_{m,-1}^0$.
The mixing of spin components could lead to the interaction between spectra 
inherited from the ones with different spin projection (at $\lambda_y=0$).
In particular, a crossing/anti-crossing behavior of two levels depends 
on their symmetry  with respect to the inversion of the $y$-axis
(see Fig.\ref{QDs}). 
Two levels with the same parity $\Psi_{1,2}=s_{1,2} \hat{M}_y\Psi_{1,2}$ 
(when $s_1=s_2$) anti-cross, while a different sign $s_1=-s_2$ 
leads to the level crossing.
Other properties of discrete levels are very similar to properties of those 
obtained at $\lambda_y=0$.

\section{Summary}
\label{sum}
Within the approach suggested by Ando (see details in \cite{Ando}), 
we solved analytically the eigenvalue problem for the effective mass 
Hamiltonian for electrons on curved surface with the spin-orbit interaction.
In particular, with the aid of transformation
(\ref{U}), we obtained explicit expressions  for a low energy spectrum 
and eigenstates of armchair carbon nanotubes. These findings have been used to analyze transport 
properties of the CNT with the curvature-induced SOC (see Eq.(\ref{lxy})) at different 
limits.

We have studied effects produced by the SOC
on the scattering of electrons at the interface introduced by the potential step
of height $V_0$. The effect of the SOC becomes
especially drastic when the height of the potential barrier can be controlled 
to reach conditions allowing to produce a spin-filter effect. At this condition
only one spin component is dominant in the transmission over the CNT.
Note that this phenomenon occurs due to the "effective magnetic field" which is
brought about by the curvature induced SOC. The SOC term $\lambda_y$ yields, however, 
a parasite loss $(\sim 10^{-3})$ of the spin-filter effect.

The gap in CNTs (due to the SOC) opens a possibility  
to create a nanotube quantum dot. In the limit of the preserved spin symmetry
we have calculated the QD eigenstates with aid of the transcendental equation.
At low energy limit the spectrum is similar to the one of the quantum well potential,
while for large energies it carries features of 
the harmonic oscillator spectrum. In such QDs  an electron
with spin up cannot scatter into state with spin down and vise versa,
without any additional mechanism.
However, the SOC term $\lambda_y$ mixes these states and  
yields the anti-crossing effect. This mechanism 
may affect the spin relaxation phenomenon in the system under consideration, 
in addition to an electron-phonon coupling mechanism \cite{Loss}. 

There was a belief that the curvature induced SOC in graphene, restricted by 
the first term, leads only to very weak back scattering \cite{Ando}. 
We have demonstrated, however, that 
the second term, ignored in a previous analysis, produces the inter-channel scattering, 
which could increase the back scattering by a few orders of
magnitude and enrich transport phenomena in carbon nanotubes.

\section*{Acknowledgments}  
K.N.P. and M.P are grateful for the congenial
hospitality at UIB and JINR. This work was supported in
part by RFBR grant 14-02-00723,
integration Grant No.29 from the Siberian Branch of the RAS
and Slovak Grant Agency VEGA grant No. 2/0037/13.

\appendix
\numberwithin{equation}{section}
\section{An eigenvalue problem for $\hat{H}'$}
\label{app1}

We suggest to use the eigenstate in the form
\begin{equation}
 F'(\theta,y)=e^{\rmi m \theta}e^{\rmi k_y y}\left(
\begin{array}{c}A\\B\\C\\D
\end{array}\right), m= \pm 1/2, \pm 3/2,....
\end{equation}
to solve the eigenvalue problem $\hat{H}'F'=E F'$
for the Hamiltonian (\ref{2}).
As a result, one obtains the Hamiltonian (\ref{Hp}). 
In virtue of the unitary transformation
\begin{eqnarray}
\label{Vtilde_av}
&{\tilde V}=
\hat{P}_{123}\hat{\tau}_x\otimes\exp\left(\rmi\frac{\pi}{4}\hat{\sigma}_x\right)=\\
&\frac{1}{\sqrt{2}}\left(
\begin{array}{cccc}
 0 & 0 & 1 & 0 \\
 1 & 0 & 0 & 0 \\
 0 & 1 & 0 & 0 \\
 0 & 0 & 0 & 1 \\
\end{array}
\right)
\left(
\begin{array}{cccc}
 0 & 0 & 1 & \rmi \\
 0 & 0 & \rmi & 1 \\
 1 & \rmi & 0 & 0 \\
 \rmi & 1 & 0 & 0 \\
\end{array}
\right)
=\frac{1}{\sqrt{2}}\left(
\begin{array}{cccc}
 1 & \rmi & 0 & 0 \\
 0 & 0 & 1 & \rmi \\
 0 & 0 & \rmi & 1 \\
 \rmi & 1 & 0 & 0 \\
\end{array}
\right)\nonumber
\end{eqnarray}
our Hamiltonian (\ref{Hp}) becomes real 
\begin{eqnarray}
\label{Hptilde_av}
&\hat{\tilde H}={\tilde V}^{-1}\hat{H}'{\tilde V}=\\
&\left(
\begin{array}{cccc}
 -\lambda_x-\lambda_y  & 0 & t_y & t_m \\
 0 & \lambda_x+\lambda_y  & t_m & -t_y \\
 t_y & t_m & \lambda_x -\lambda_y & 0 \\
 t_m & -t_y & 0 & -\lambda_x+\lambda_y  \\
\end{array}
\right)\nonumber
\end{eqnarray}
with notations (\ref{abbrev}).
The eigenvalues of the Hamiltonian (\ref{Hptilde_av}) are
\begin{eqnarray}
\label{Epm_av}
&E=\pm E_{m,s}\,,\\
&E_{m,s} = \sqrt{t_m^2+t_y^2+\lambda_y^2+\lambda_x^2+2 D_{m,s}}\,,\quad s=\pm 1\,,\nonumber\\
&D_{m,s} = s \sqrt{\lambda_x^2\left(t_m^2+\lambda_y^2\right)+t_y^2\lambda_y^2}.\nonumber
\end{eqnarray}
Eigenvectors have rather simple form
\begin{eqnarray}
\label{Vec_av}
&&{\cal V}_{m,s}=\frac{1}{N_{m,s}}\Bigg{(}
\frac{-D_{m,s}+\lambda_y(E-\lambda_y)}{t_m(\lambda_x+\lambda_y)},\\
&&\frac{-t_y(\lambda_x-\lambda_y)}{-D_{m,s}+\lambda_x(E-\lambda_x)},
\frac{t_y}{t_m}\frac{(-D_{m,s}+\lambda_y(E-\lambda_y))}{(-D_{m,s}+\lambda_x(E-\lambda_x))},
1
\Bigg{)}^T,\nonumber
\end{eqnarray}
where the values $E=\pm E_{m,s}$, $D_{m,s}$ are defined by Eqs.(\ref{Epm_av}),
and the norm $N_{m,s}$ is
\begin{equation}
\begin{array}{rcl}
N_{m,s}&=&\sqrt{({\cal F}_1+{\cal F}_2+{\cal F}_3)/{\cal M}+1},\\
{\cal F}_1&=&\left(D_{m,s}-\lambda_y(E-\lambda_y)\right)^2\left(D_{m,s}-\lambda_x(E-\lambda_x)\right)^2,\\
{\cal F}_2&=&t_m^2 t_y^2(\lambda_x^2-\lambda_y^2)^2,\\
{\cal F}_3&=&t_y^2(\lambda_x+\lambda_y)^2\left(D_{m,s}-\lambda_y(E-\lambda_y)\right)^2,\\
{\cal M}&=&t_m^2 (\lambda_x+\lambda_y)^2 (D_{m,s}-\lambda_x(E-\lambda_x))^2.
\end{array}
\end{equation}

The expectation value of the current for eigenspinors (\ref{Vec_av}) could be calculated 
with the aid of the definitions (\ref{J}) and the 
transformation (\ref{Vtilde_av}). As a result, we obtain
\begin{eqnarray}
\label{J_av}
&&\langle\hat{j}_y\rangle_{m,s}=\frac{\gamma t_y}{E}
\left(1+\frac{\lambda_y^2}{D_{m,s}}\right),\nonumber\\
&&\langle\hat{j}_\theta\rangle_{m,s}=\frac{\gamma t_m}{E}
\left(1+\frac{\lambda_x^2}{D_{m,s}}\right).
\end{eqnarray}

To solve a scattering problem we need the eigenspinors for a fixed value of 
the energy $E$. Using the dispersion relation (\ref{Epm_av}), we obtain
\begin{equation}
\label{kpm_av}
k_{m,s}=\frac{1}{\gamma}\sqrt{E^2+\lambda_y^2-t_m^2-
\lambda_x^2-2 s\sqrt{\lambda_y^2E^2+t_m^2(\lambda_x^2-\lambda_y^2)}}\,.
\end{equation}
Substituting the value of $k_{m,s}$ from Eq. (\ref{kpm_av}),
the spinor (\ref{Vec_av}) becomes ${\cal V}_{m,s}(t_y=\gamma k_{m,s})$ with
the parameter $D_{m,s}$ simplified to the form
\begin{equation}
D_{m,s}=s\sqrt{\lambda_y^2E^2+t_m^2(\lambda_x^2-\lambda_y^2)}-\lambda_y^2.
\end{equation}

\end{document}